\documentclass[9pt]{article}
\usepackage{spconf,amsmath,graphicx}
\usepackage{amsfonts,amssymb}
\usepackage{booktabs}

\title{INCREASING COMPACTNESS OF DEEP LEARNING BASED SPEECH ENHANCEMENT MODELS WITH PARAMETER PRUNING AND QUANTIZATION TECHNIQUES}
\name{Jyun-Yi Wu$^\ast$, Cheng Yu$^\ast$, Szu-Wei Fu$^\dagger$, Chih-Ting Liu$^\ast$, Shao-Yi Chien$^\ast$, Yu Tsao$^\dagger$\thanks{Thanks to IOX center, NTU.}}
\address{$^\ast$Graduate Institute of Electronic Engineering and Department of Electrical Engineering,\\ National Taiwan University, Taiwan  \\ $^\dagger$Research Center for Information Technology Innovation, Academia Sinica, Taiwan}
\begin{document}
%
\maketitle
\begin{abstract}
Most recent studies on deep learning based speech enhancement (SE) focused on improving denoising performance.  
However, successful SE applications require striking a desirable balance between denoising performance and computational cost in real scenarios.  
In this study, we propose a novel parameter pruning (PP) technique, which removes redundant channels in a neural network. 
In addition, a parameter quantization (PQ) technique was applied to reduce the size of a neural network by representing weights with fewer cluster centroids.
Because the techniques are derived based on different concepts, the PP and PQ can be integrated to provide even more compact SE models.
The experimental results show that the PP and PQ techniques produce a compacted SE model with a size of only 10.03 $\%$ compared to that of the original model, resulting in minor performance losses of 1.43$\%$ (from 0.70 to 0.69) for STOI and 3.24$\%$ (from 1.85 to 1.79) for PESQ.  
The promising results suggest that the PP and PQ techniques can be used in an SE system in devices with limited storage and computation resources.
\end{abstract}

\section{Introduction}
\label{sec:intro}

The goal of speech enchantment (SE) is to generate enhanced speech with better quality and intelligibility over the original noisy speech. Many SE methods have been proposed in the past. One class of approaches directly deducts the estimated noise components from noisy speech in the spectral domain; notable examples include spectral subtraction \cite{SPECTRAL} and its extensions. Another class of approaches considers the characteristics of speech and noise signals during the design of a gain function, which is used to filter out the noise components; well-known examples include the Wiener filter \cite{Wiener}, minimum mean-square-error (MMSE) \cite{extension1}, and maximum-likelihood spectral amplitude (MLSA) \cite{MLSA2} algorithms. These traditional approaches perform well when the assumed properties of speech and noise signals are maintained, while the performance degrades notably when dealing with non-stationary noises or operating with very low signal-to-noise ratio (SNR). 
\par
Recently, deep learning algorithms have been successfully introduced to the SE field \cite{SUPERVISEDSS}. Generally speaking, a deep-learning model is used as a mapping function with the aim of transforming noisy speech into clean speech. Notable approaches include the deep denoising auto-encoder (DDAE) \cite{DDAE1}, deep feedforward neural network \cite{FFNN2}, convolutional  neural network (CNN) \cite{CNN1}  and long short-term memory model (LSTM) \cite{LSTM}; all of these models have shown promising results for transforming noisy spectral features into clean ones. More recently, several studies proposed the use of convolutional structures for speech and audio signal analyses and reconstruction \cite{FCN, FCN-Fu, Time, tasnet, Multi}, and thus the SE tasks can be directly carried out in the time domain. 

Numerous studies have confirmed the outstanding denoising capability of deep learning-based methods, especially under more challenging conditions (e.g., non-stationary noises and low SNR conditions). However, a notable disadvantage of deep learning-based solutions is the requirement of large storage space for the SE models and high online computational costs, which makes them difficult to implement in a device with limited resources. In this study, we propose two techniques, namely parameter pruning (PP) and parameter quantization (PQ), to increase the compactness of deep learning-based SE models. The PP technique removes redundant channels and the PQ technique groups and represents similar weights using a cluster centroid. To evaluate the effectiveness of these two techniques, we used the TIMIT database  \cite{TIMIT} with several noise sources. In this study, we focus on the waveform mapping based SE method using the Fully Convolutional Neural Network (FCN) model. The experimental results show that both PP and PQ techniques can effectively improve the model compactness with modest degradations in quality and ineligibility performance. 
\par
The rest of this paper is organized as follows. Related research is reviewed in Section 2. Section 3 introduces the proposed techniques. Section 4 presents the experimental setup and results. Our concluding remarks are stated in Section 5.
\vspace{-0.1cm}
\section{Related Research}
\label{sec:part2}

As mentioned earlier, we focus our attention on waveform mapping-based SE using the FCN model. The FCN model is a specialized CNN model that consists of only convolutional layers. In our previous studies, we showed that the FCN model can be used to directly map a noisy speech waveform to a clean waveform  \cite{FCN, FCN-Fu}. 
There are two advantages of the waveform mapping-based SE process. First, possible distortions caused by imperfect phase information can be alleviated; second, the computational cost for converting a waveform to frame-based spectral features can be reduced.


\par
Many algorithms have been derived to increase the compactness of neural network models, such as pruning, sparse constraints, and quantization. Pruning algorithms are de- signed to reduce the network complexity and to address overfitting \cite{Pruning1-Han}. By determining a threshold, any weight values lower than the threshold are removed from the model, thus reducing the total number of weights. Another class of approaches builds compact models by applying sparse constraints to reduce trivial filters in the models \cite{Sparsity5-Liu}. On the other hand, quantization algorithms compress the size of the original network by reducing the number of bits required to represent each weight \cite{log1-Miyashita}. Han et al. \cite{kmeans2-Han} applied a k-means scalar quantization to the parameter values. These quantization methods significantly reduced memory usage with a modest loss in recognition accuracy.

Based on our literature survey, only few studies have investigated potential approaches to increase the compactness of SE models. Sun and Li proposed using a quantization technique to increase the compactness of an SE model \cite{sun2017optimization}. Ko et al. investigated the correlation of precision scaling and neuron numbers in an SE model \cite{ko2017precision}.  In \cite{hsu2018study}, a two-stage quantization approach was derived to optimally reduce the number of bits when the model parameters are encoded in floating point representation. In the present study, the proposed PP technique adopted a different and novel concept that directly removes redundant channels to form a compact FCN model. The size of this model is then reduced further with the PQ technique.  
\label{sec:part3}

\vspace{-0.3cm}
\section{The proposed PP and PQ Techniques}
This section introduces the proposed PP and PQ techniques, as well as the integration of these two techniques.
\vspace{-0.3cm}
\subsection{The Parameter Pruning (PP) Technique}
\subsubsection{FCN-based Waveform Mapping}
\label{ssec:subhead}
\begin{figure}[t]
\begin{minipage}[t]{1\linewidth}
  \centering
  \centerline{\includegraphics[height=6.3cm,width=8cm]{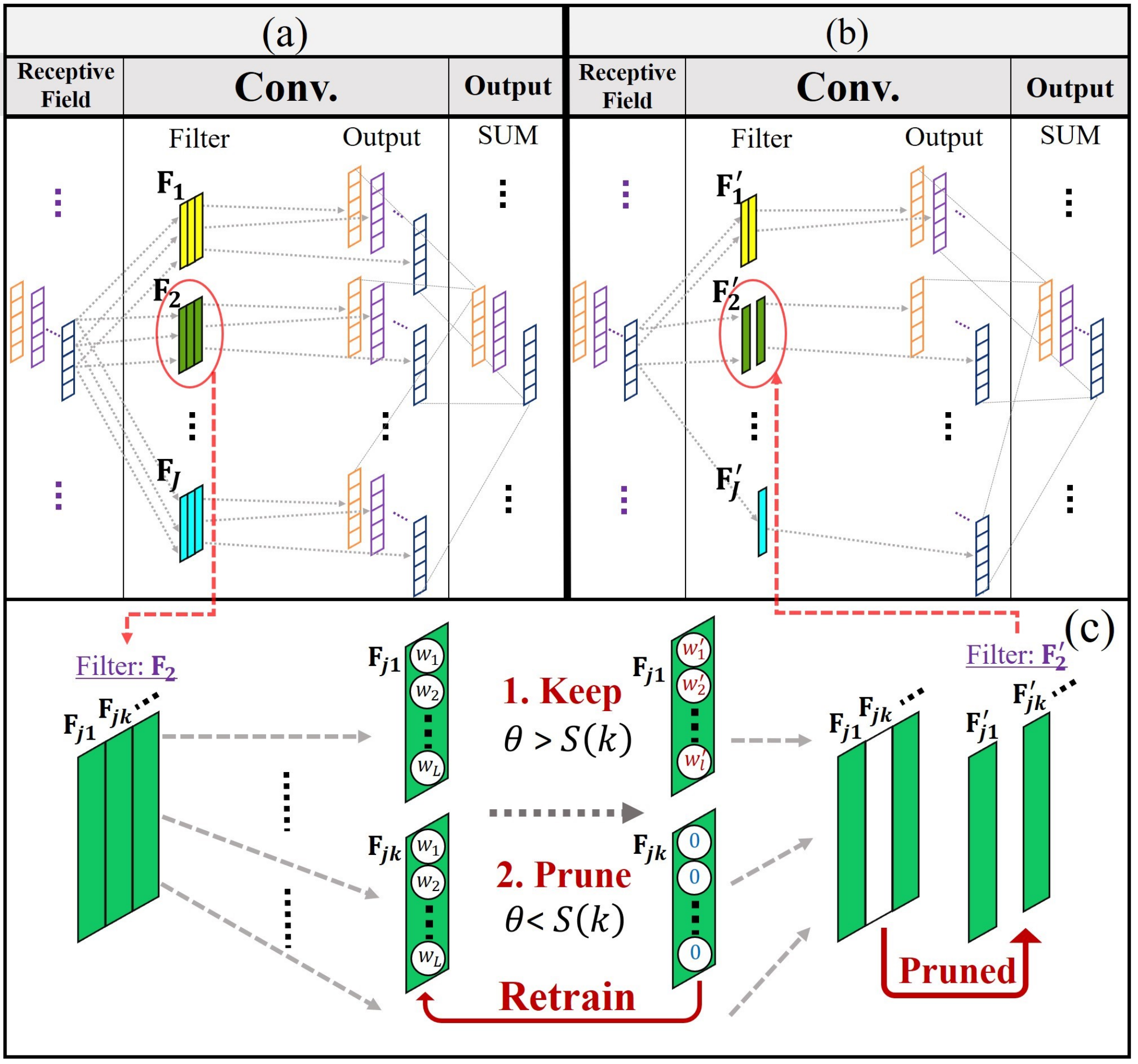}}
\end{minipage}
\vspace{-0.7cm}
\caption{The PP process: (a) Original model; (b) pruned model, and (c) the pruning and retraining process.}
\label{fig:filter}
\end{figure}

Figure \ref{fig:filter}(a) shows the process of the waveform mapping based on the FCN model. In the figure, we have $J$ filters: $\{ $ \textbf{F}$_{1}$, \textbf{F}$_{2}$, ..., \textbf{F}$_{J}$$\}$; \textbf{F}$_{j}$ $\in \mathbb{R}^{ L \times I}$ is the $j$-th filter, and \textbf{F}$_{ji}$ $\in \mathbb{R}^{ L \times 1}$ is the $i$-th channel of \textbf{F}$_{j}$. \textbf{F}$_{ji}$ $\ = \ ( $ $w$$_{1}$, $w$$_{2}$, ..., $w$$_{L}$  $)_{L \times 1}$ where $w_{l}$ is the channel weight. Assume that the receptive field and output sample of filter \textbf{F}$_{j}$ is \textbf{R}(t) $\in \mathbb{R}^{ L \times I}$ and \textbf{y$_{j}$}(t), repectively. The resulting convolution operation is:  
\begin{equation}
\textbf{y$_{j}$}(t) \ = \ \sum_{i=1}^{I} \textbf{F}_{ji}^{T} \textbf{R$_{i}$}(t). 
\end{equation}
\label{sssec:subsubhead}
\label{ssec:subhead}
\vspace{-0.7cm}
\subsubsection{Definition of Sparsity}
We define the redundancy criterion with sparsity of each channel in a filter. For filter \textbf{F}$_{j}$ in an arbitrary layer of the FCN, we first compute the mean absolute value of all filter weights:

\begin{equation}
M_{F_j} \ = \ \frac{\sum_{I}(\sum_{L}|w|)}{I \times L},
\label{eq:mean}
\end{equation}
\par\noindent
where $I$ and $L$ are the total number of channels in a filter and number of weights in a channel, respectively, and $w$ is a weight parameter. The sparsity of the $i$-th channel in a filter \textbf{F}$_{j}$ can then be defined as:
\begin{gather}
S(i) \ = \ \frac{\sum_{l=1}^{L}\sigma(w_l)}{ L}, \\
\sigma(X)=
\begin{cases} 
1,  &  \ \mbox{if }|X|<M{_F{_j}} \\
0,  &  \ \mbox{otherwise}.
\end{cases}
\end{gather}
\par\noindent
When $S(i)$ is close to $1$, most of weights in a channel are smaller than $M_{F_j}$, and the channel is considered more redundant.
\vspace{-0.3cm}
\subsubsection{Channel Pruning}
In our proposed parameter pruning (PP) technique, the pruning mechanism contains a retraining step. As shown in Fig.\ref{fig:filter}.(c), if the sparsity $S(k)$ in some channels  \textbf{F}$_{jk}$ are larger than a predefined threshold value $\theta$, the weights within the channel \textbf{F}$_{jk}$ will be set to zero. Next, we retrain the model. After several iterations, we then remove \textbf{F}$_{jk}$, as shown in Fig.\ref{fig:filter}.(b), we can then obtain \textbf{F}$^{'}$ as the channel-pruned filters. Because \textbf{F}$^{'}_{ji}$ is reduced, \textbf{R}$^{'}_{i}$(t) is reduced accordingly after the PP process. Finally, we can compute the output as follows:
\begin{equation}
\textbf{y$_j^{'}$}(t) \ = \ \sum_{i=1}^{I-K}\textbf{F}_{ji}^{'T}\textbf{R$_{i}^{'}$}(t),
\end{equation}
where $K$ is the number of pruned channels. This PP technique ensures that the compacted model remains stable after the pruning steps, while the retraining steps make models adjustable to the zero-weighted channels. We believe that our approach, unlike other pruning methods that directly remove filters, can effectively prevent severe performance drops.

\label{sec:part3}
\subsection{The Parameter Quantization (PQ) Technique}
\label{ssec:subhead}
\begin{figure}[h]
\begin{minipage}[t]{1\linewidth}
  \centering
  \centerline{\includegraphics[width=8cm]{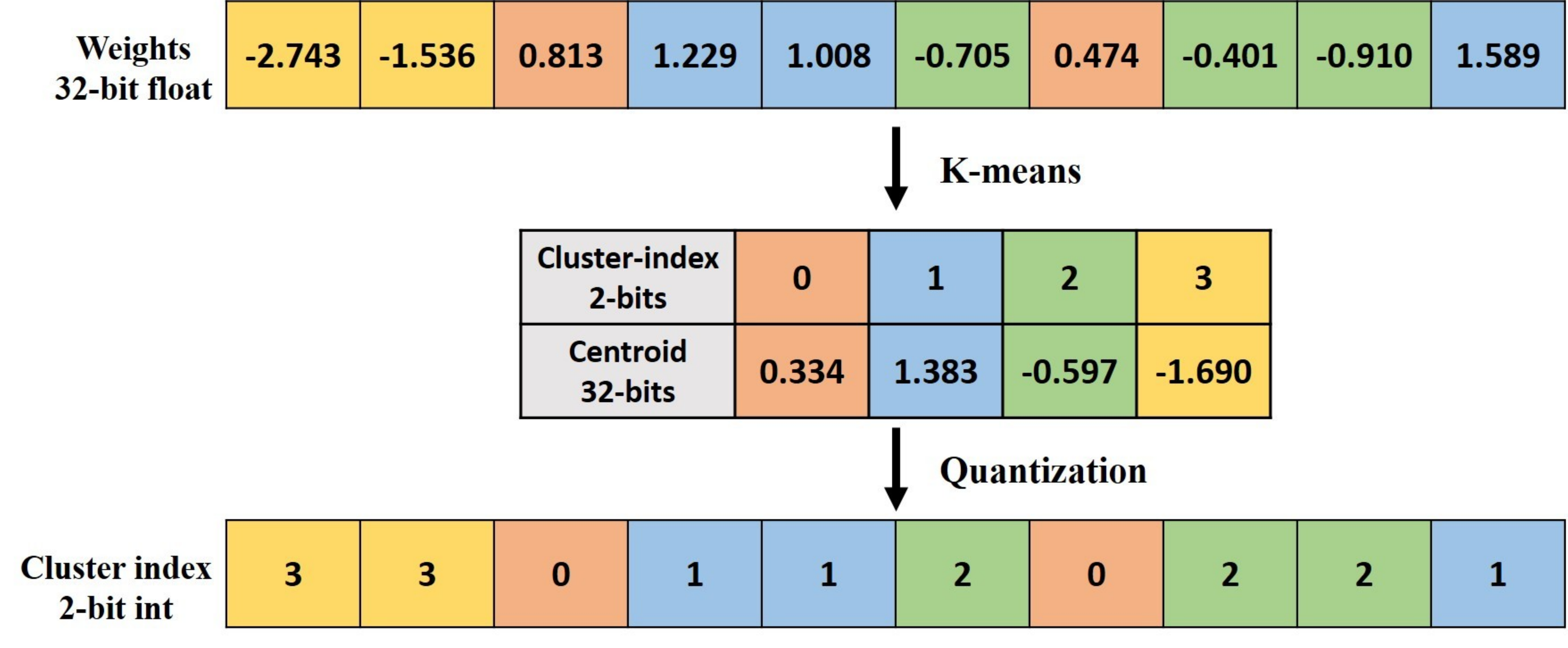}}
\end{minipage}
\vspace{-0.7cm}
\caption{Example of the PQ technique.}
\label{fig:Kmeans}
\end{figure}

In this study, the parameter quantization is carried out based on the k-means algorithm. By applying the k-means quantization, the parameters in a neural network model are grouped into several clusters, where each cluster of parameters shares a centroid value. Fig. \ref{fig:Kmeans} shows an example of the k-means-based PQ process. In this figure, each weight parameter in the original model is represented by a 32-bits floating point number. By applying the k-means algorithm with k=4, we can obtain a look-up table with 4 cluster centroids. Each weight in the model is then denoted with a cluster index that is linked to the corresponding cluster centroid. Therefore, the 10 weights (each represented as a 32-bit floating point number) in the original model can be represented with 4 cluster indices and 4 centroids. The corresponding compression rate is:
$(10 \ * \ 32)/(4*32 \ + \ 2*10) \ = \ 2.16$.
\vspace{-0.3cm}
\vspace{-0.5cm}

\subsection{The Integration of PP and PQ}
\label{ssec:subhead}
\begin{figure}[htb]
\begin{minipage}[t]{1\linewidth}
  \centering
  \centerline{\includegraphics[width=8.5cm]{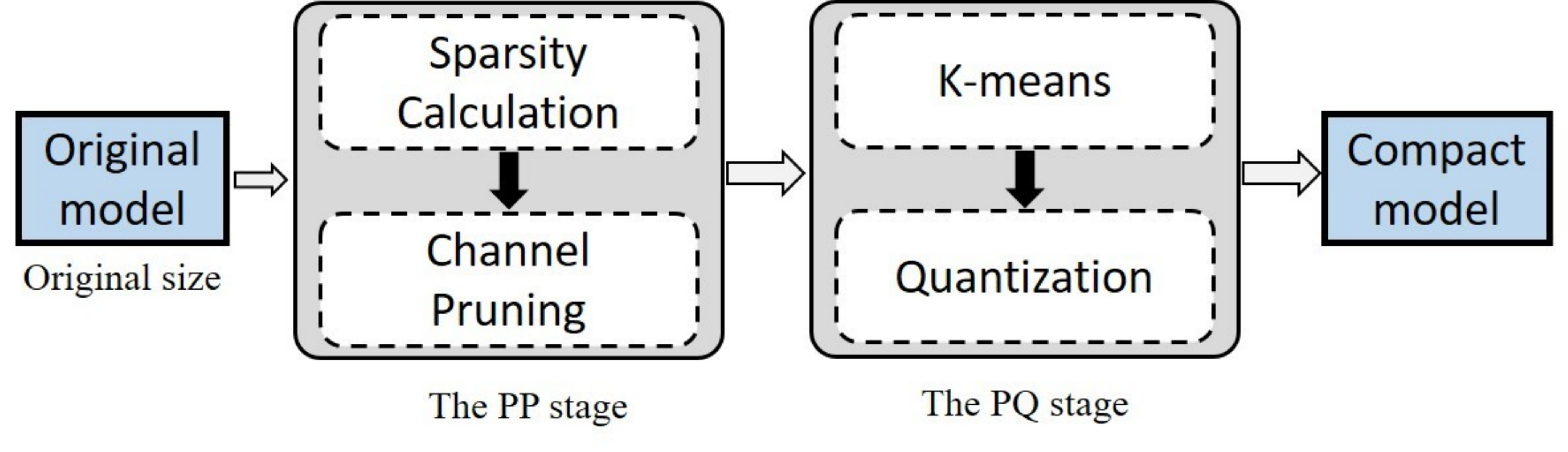}}
\end{minipage}
\vspace{-0.7cm}
\caption{Integration system of the PP and PQ techniques.}
\label{fig:process}
\end{figure}

Although both methods aim to increase the model compactness, the PP and PQ techniques are derived based on different concepts. In this section, we investigate the compatibility of these two techniques. 
Fig. \ref{fig:process} shows the proposed integration system. PP is applied to remove redundant channels and establish a compact SE model. PQ is subsequently used to further reduce the model size. 

\section{Experiments}
\label{sec:part4}
In this section, we first introduce the experimental setup and then demonstrate the experimental results.

\subsection{Experimental Setup}
\label{ssec:subhead}
The TIMIT corpus was used to prepare the training and test sets. All 4620 utterances in the TIMIT training set were selected as training data. These utterances were corrupted with five noise types (Babble, Car, Jackhammer, Pink, and Street) at five SNR levels (-10 dB, -5 dB, 0 dB, 5 dB, and 10 dB). 100 utterances were randomly selected from the TIMIT testing set as the testing data. These utterances were artificially corrupted with three noise types (Babycry, White, and Engine) at four SNR levels (-12 dB, -6 dB, 0 dB, and 6 dB). Note that we intentionally designed mismatched noise types and SNR levels for training and testing conditions in order to simulate a more realistic scenario.
We evaluated the PP and PQ techniques using two standardized metrics: perceptual evaluation of speech quality (PESQ) \cite{PESQ} and short-time objective intelligibility (STOI) \cite{STOI}. PESQ measures the quality of the processed speech by assigning a score ranging from −0.5 to 4.5; a higher PESQ score denotes better speech quality. STOI measures speech intelligibility by assigning a score ranging from 0 to 1; a higher STOI score denotes better intelligibility. 

\subsection{Experimental Results}
In this section, we present the PESQ and STOI results produced with PQ, PP, and the integrated system. 
\vspace{-0.3cm}
\subsubsection{Parameter Quantization (PQ)}
\label{ssec:subhead}
\begin{figure}[h]
\begin{minipage}[b]{0.48\linewidth}
  \centering
  \centerline{\includegraphics[width=4.0cm]{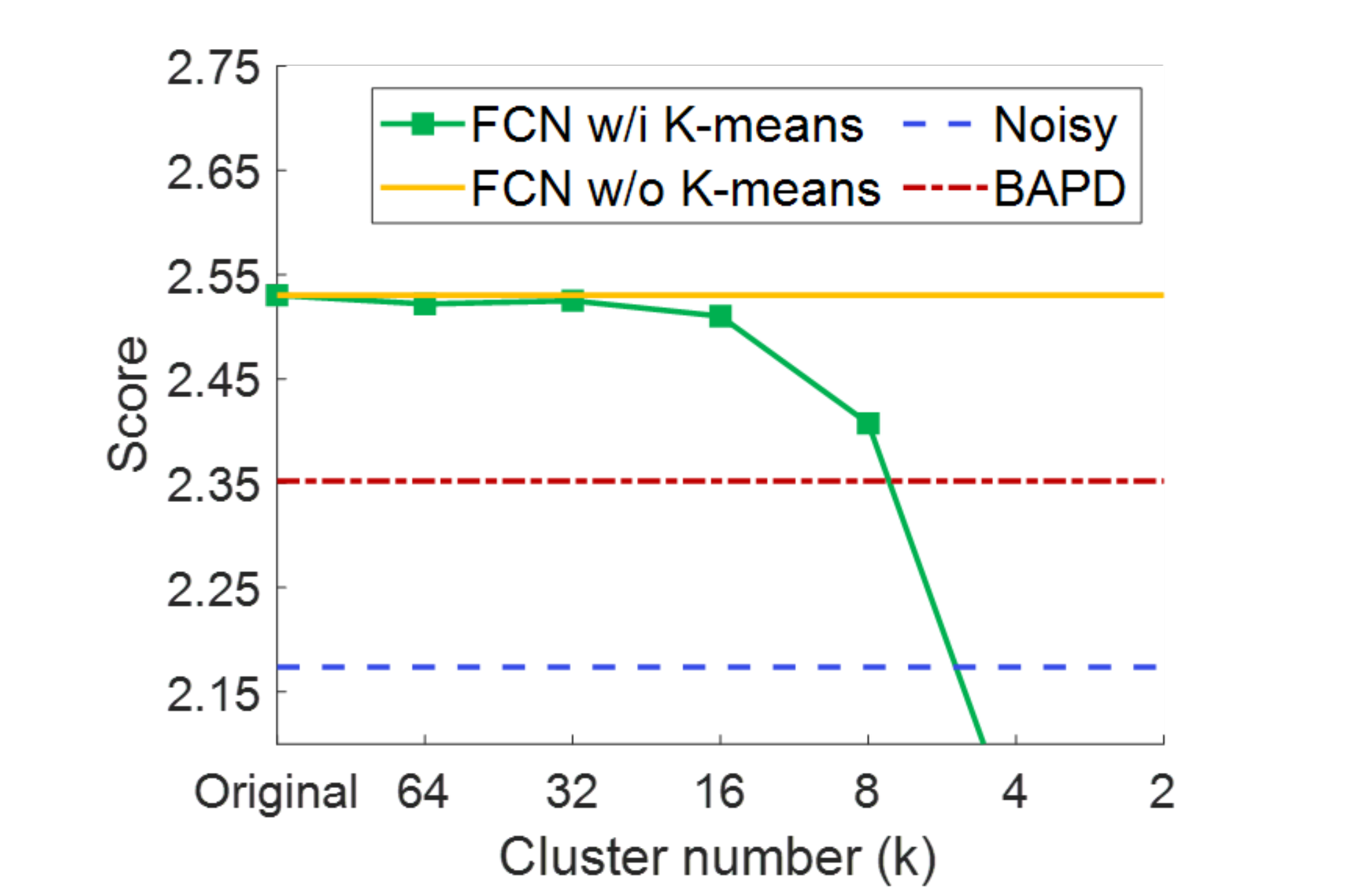}}
  \centerline{(a) PESQ}\medskip
  \label{fig:k.means.result.pesq}
\end{minipage}
\hfill
\begin{minipage}[b]{0.48\linewidth}
  \centering
  \centerline{\includegraphics[width=4.0cm]{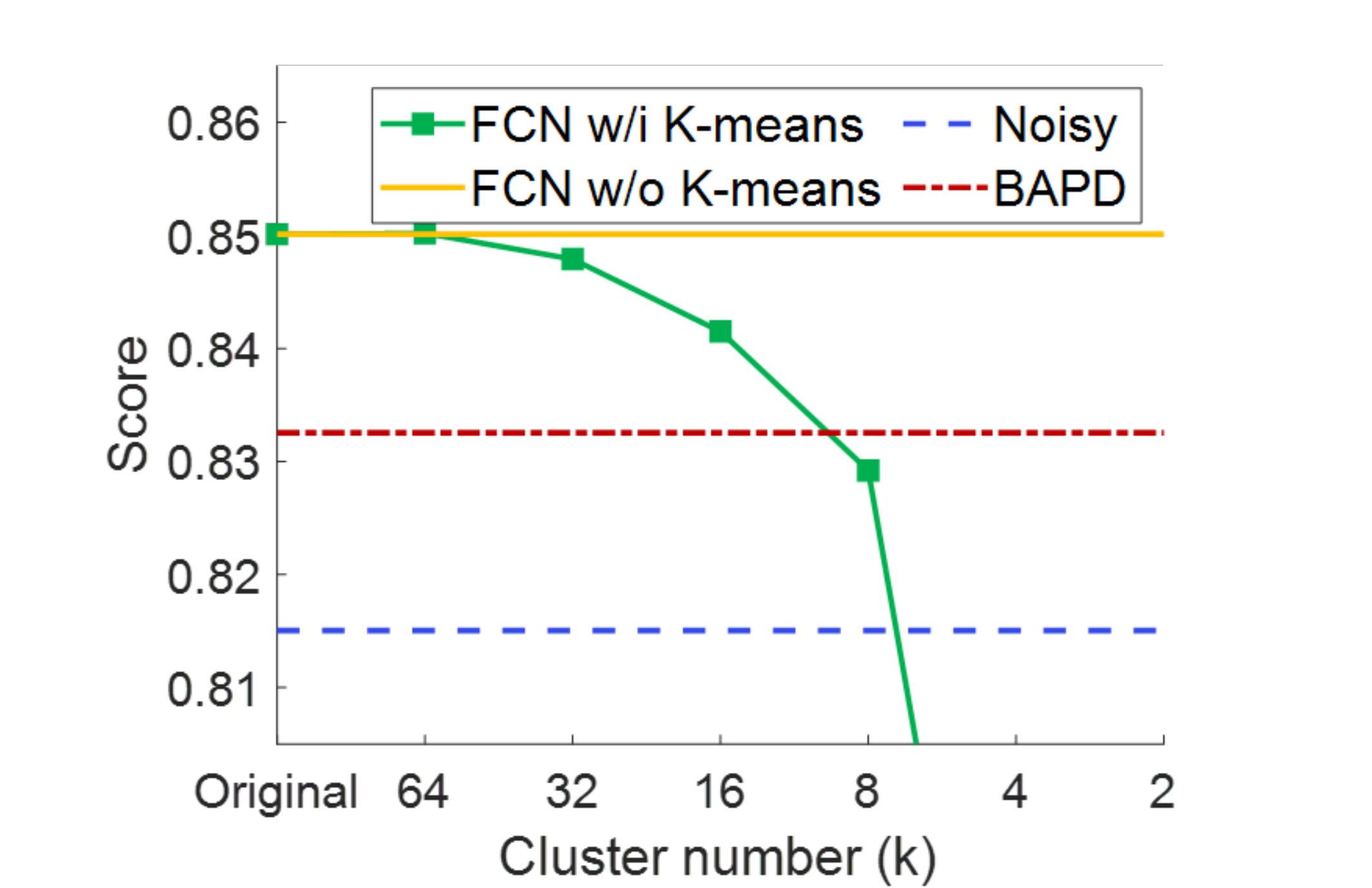}}
  \centerline{(b) STOI}\medskip
  \label{fig:k.means.result.stoi}
\end{minipage}
\vspace{-0.4cm}
\caption{The average PESQ and STOI scores yielded from the PQ technique with different numbers of clusters. BAPD denotes the bound for acceptable performance drop.}
\label{fig:pqresult}
\end{figure}

Regarding the PQ technique, we set the number of clusters k to 2, 4, 8, 16, 32, and 64, and the corresponding PESQ and STOI results are shown in Fig. \ref{fig:pqresult} (a) and (b), respectively. It is clear that the scores of both PESQ and STOI decrease when the cluster number
k is reduced. In practical SE applications, we must consider performance and computation simultaneously. Thus, we may first define a bound for acceptable performance drop (BAPD) and continue reducing the cluster number until the evaluation scores are lower than a defined bound. In this experiment, we consider this BAPD to be the mean score of the results produced with the original SE model and that of noisy speech. Using Fig.\ref{fig:pqresult}(a) as an example, the PESQ scores for noisy speech and FCN without pruning are 1.64 and 1.85, respectively. BAPD is then defined as (1.64+1.85)/2 = 1.75.  
It is clear that this BAPD should be determined based on the target task. Here, we used a remarkably simple method for determining the BAPD value by providing an example. 
Figs. \ref{fig:pqresult}(a) and (b) show that the PESQ and STOI scores are similar with reduced BAPD when k$>$4.

\vspace{-0.8cm}
\subsubsection{Parameter Pruning (PP)}
\label{ssec:subhead}
\begin{figure}[h]
\begin{minipage}[b]{.48\linewidth}
  \centering
  \centerline{\includegraphics[width=4.0cm]{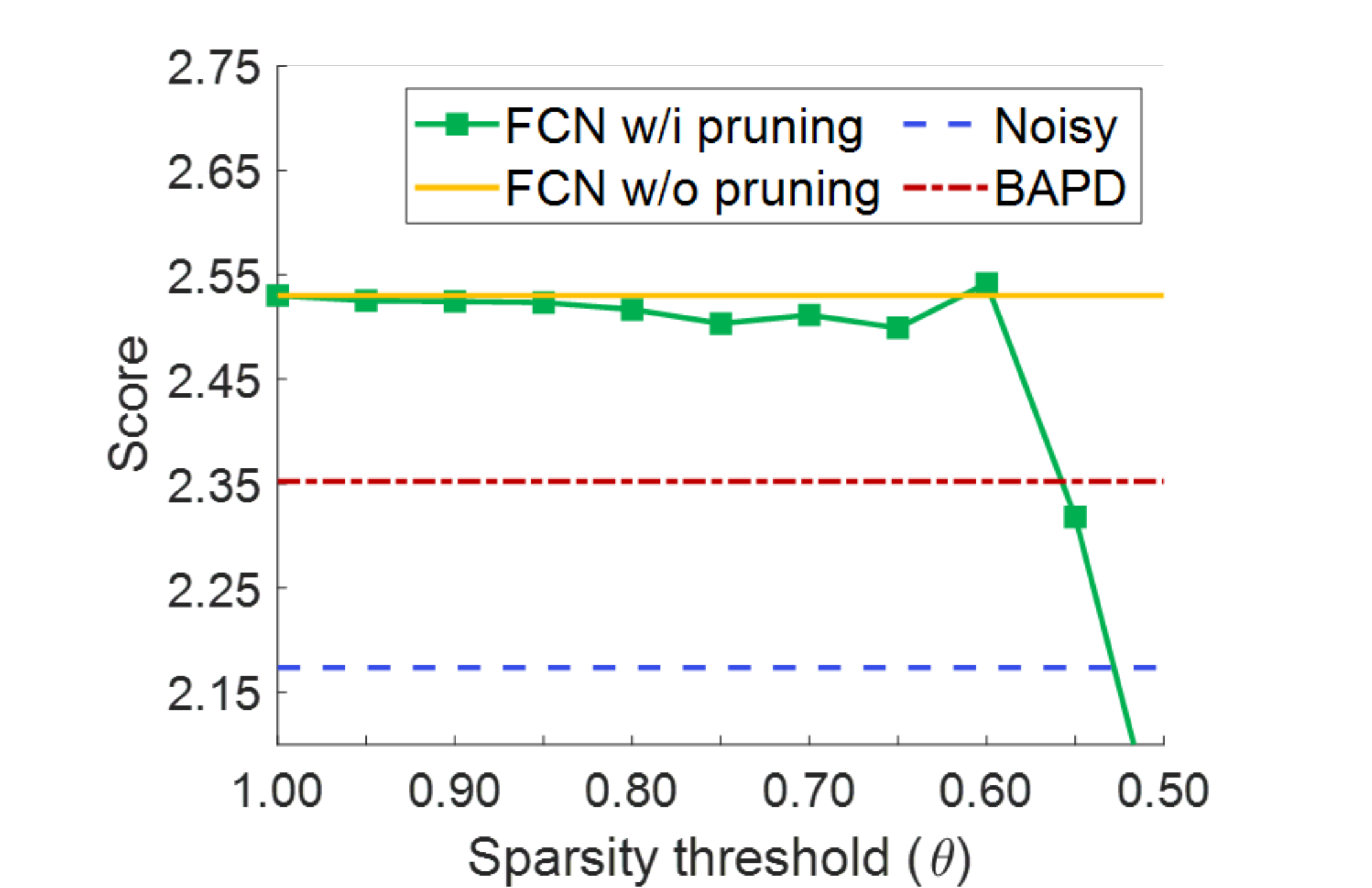}}
  \centerline{(a) PESQ}\medskip
  \label{fig:pruning.result.pesq}
\end{minipage}
\hfill
\begin{minipage}[b]{0.48\linewidth}
  \centering
  \centerline{\includegraphics[width=4.0cm]{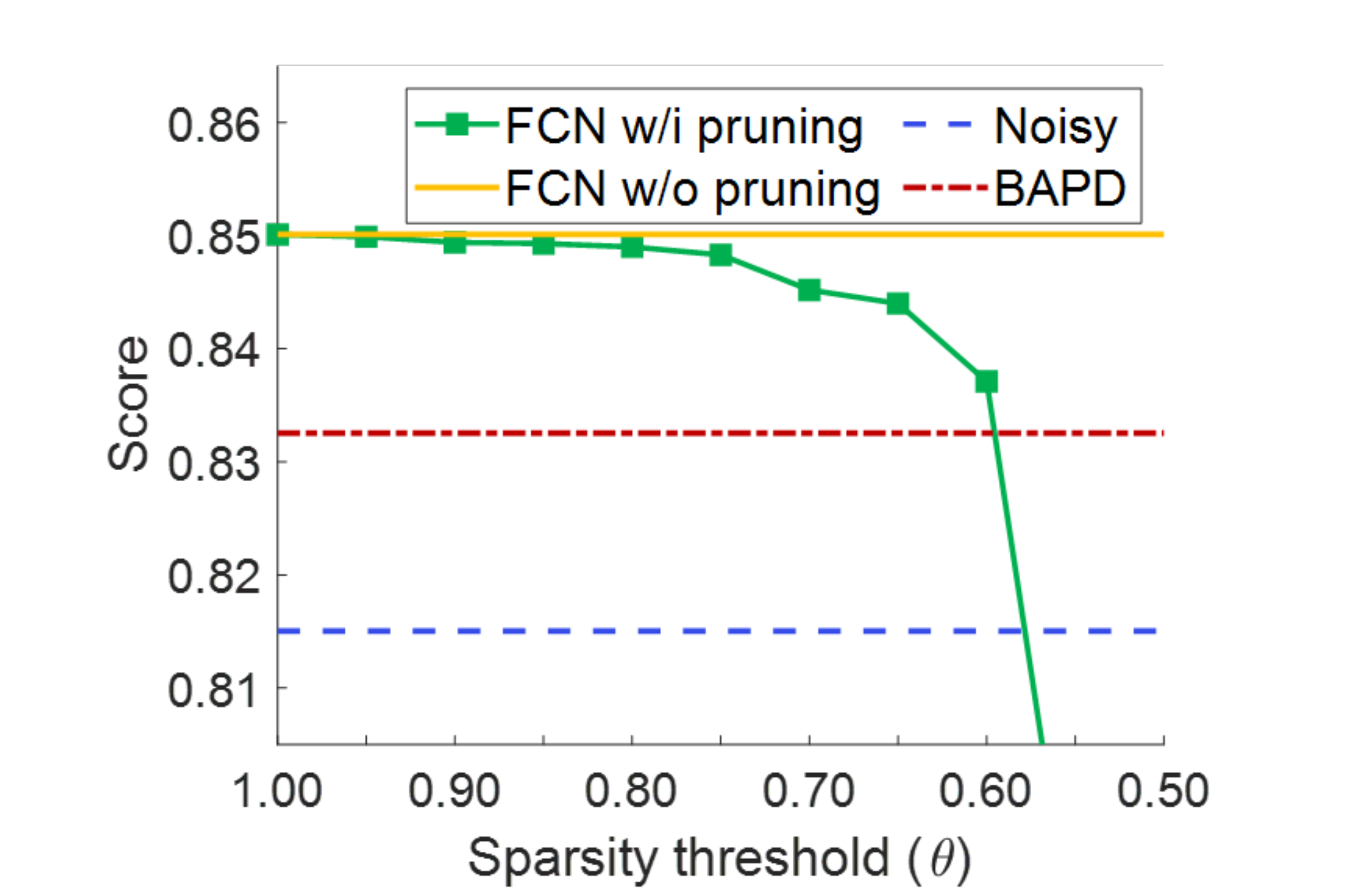}}
  \centerline{(b) STOI}\medskip
  \label{fig:pruning.result.stoi}
\end{minipage}
\vspace{-0.4cm}
\caption{The average PESQ \& STOI scores yield by the PP technique with different sparsity threshold value. BAPD denotes the bound for acceptable performance drop.}
\label{fig:ppresult}
\end{figure}

While implementing the PP technique, we gradually reduced the sparsity threshold from 1 (i.e., without conducting PP) to 0.60 with a step size of 0.05. We retrained the model after each sparsity threshold reduction. The PESQ and STOI results are shown in Figs. \ref{fig:ppresult} (a) and (b), respectively. These results show that both PESQ and STOI scores significantly dropped when the sparsity threshold decreased from 0.7 to 0.65. In table 1, we listed the correlation between the sparsity threshold and the removal ratio in the SE model. The results in Table 1 show that the corresponding removal ratio is 19.8$\%$ when the sparsity threshold is set to 0.7.

\vspace{-0.3cm}
\begin{table}[h]
		\caption{Correlation between sparsity threshold and removal ratio, as well as the number of remaining parameters in the SE model. } 
		\centering 
		\begin{tabular}{c c c } 
			\hline\hline 
			Sparsity threshold & Removal ratio & Remaining parameters  \\
			\hline
			1.00               & 0.00\%             & 300,300             \\

0.75               & 14.0\%             & 258,225             \\
0.70               & 19.8\%             & 240,900             \\
0.65               & 27.1\%             & 218,900             \\
0.60               & 30.1\%             & 209,770             \\
			\hline\hline 
		\end{tabular}
		\label{tab:pp}
	\end{table}

\vspace{-0.3cm}
\subsubsection{The Integration of PP and PQ}
\label{ssec:subhead}
\begin{figure}[htb]
\begin{minipage}[b]{0.48\linewidth}
  \centering
  \centerline{\includegraphics[width=4.0cm]{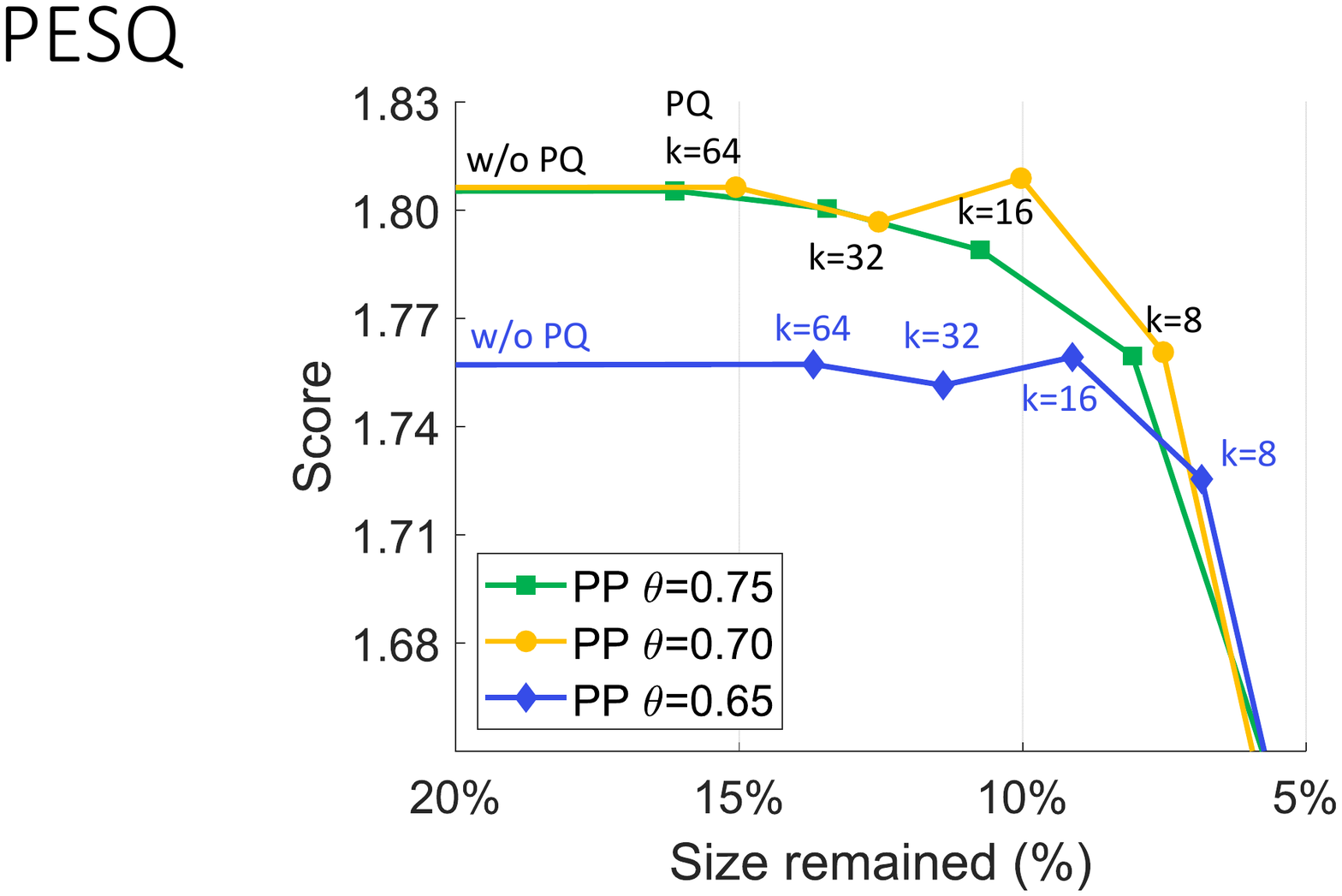}}
  \centerline{(a) PESQ}\medskip
  \label{fig:pruning.result.pesq}
\end{minipage}
\hfill
\begin{minipage}[b]{0.48\linewidth}
  \centering
  \centerline{\includegraphics[width=4.0cm]{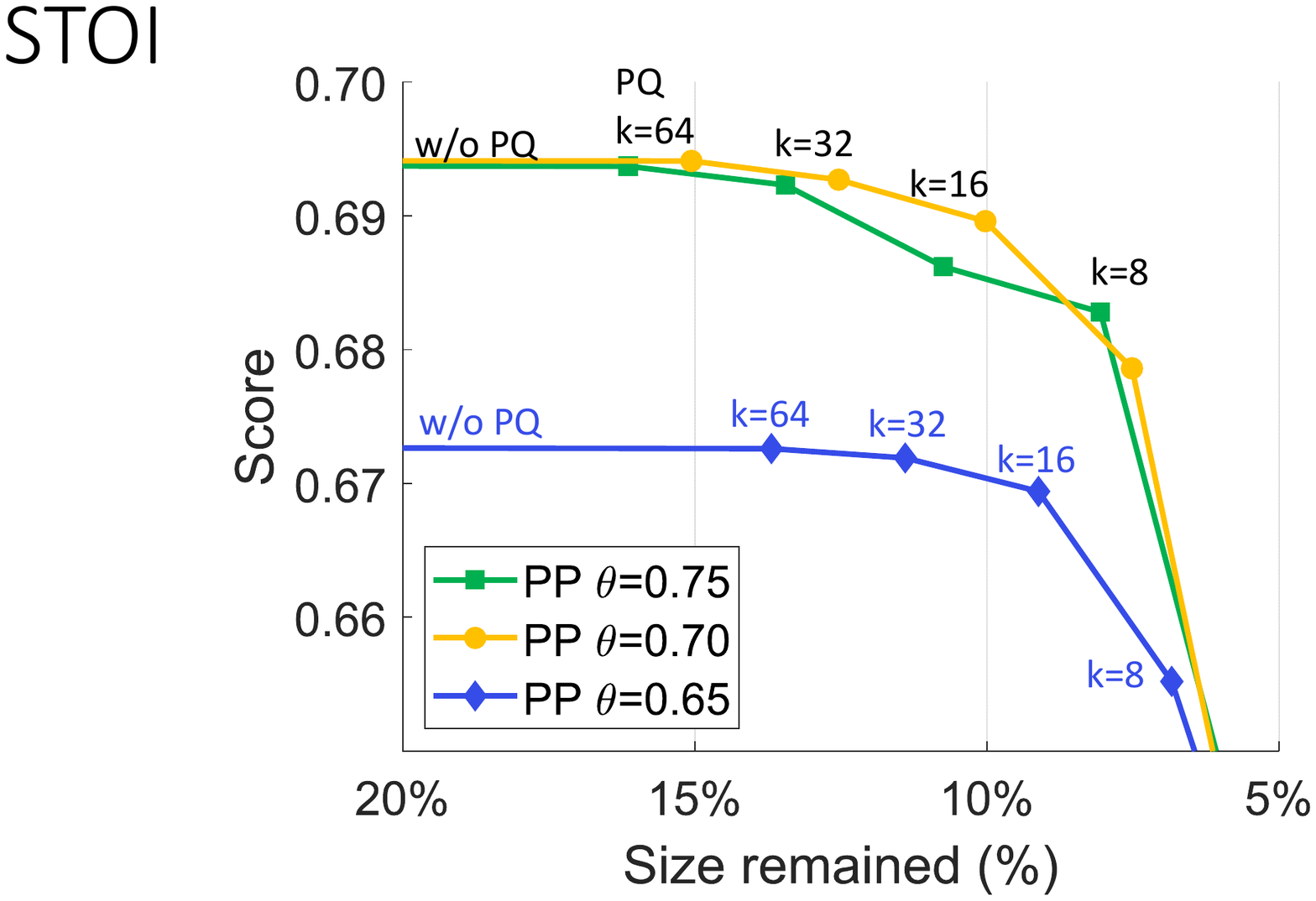}}
  \centerline{(b) STOI}\medskip
  \label{fig:pruning.result.stoi}
\end{minipage}
\vspace{-0.4cm}
\caption{The average of PESQ \& STOI results achieved by the integration of PP and PQ techniques.}
\label{fig:exp.combine.result}
\end{figure}

Finally, we investigated whether integration of the PP and PQ techniques can provide an even more compact SE model. Based on our preliminary experiments, a more effective integration order is to use PP followed by PQ. From the results in Figs. \ref{fig:pqresult} and \ref{fig:ppresult}, setting sparsity threshold $\theta$=0.70 provides reasonably satisfactory performance. Therefore, we tested $\theta$ of 0.65, 0.70, and 0.75 with the PP technique while varying the number of clusters. 
The results are shown in Fig. \ref{fig:exp.combine.result}. 
We first note that the systems with $\theta$=0.70 and 0.75 provide similar performance; both value of $\theta$ notably outperform the case with $\theta$=0.65. Moreover, the systems with $\theta$=0.70 and 0.75 both suffered considerable performance drops when k=8. 
The results in Fig. \ref{fig:exp.combine.result} show that the system with $\theta$=0.70 and k=16 provides the best performance: The size of the compacted SE model is only 10.03 $\%$ as compared to that of the original model, where STOI reduces by 1.43\% (from 0.70 to 0.69) and PESQ reduces by 3.24\% (from 1.85 to 1.79).  
\vspace{-0.5cm}
\section{Conclusion}

\label{sec:part5}
We propose utilizing the PP and PQ techniques to increase the compactness of the FCN model for an SE task. The main contribution of this study is two-fold. First, to the best of our knowledge, the PP technique is the first technique that directly removes redundant channels in the FCN model. Second, we have shown that applying PP, PQ, and an integration of PP and PQ effectively reduces the model size with only modest performance drops. The results suggest that the use of the proposed PP and PQ techniques allow an SE system with a compact neural network model to be installed in embedded devices that have lower computational capabilities. Please also note that although compression techniques for deep-learning models for pattern recognition (classification) tasks have been popularly studied, there are only few works researching model compression for signal generation (regression) tasks. Because of different output formats, the effects of model compression on regression tasks should be very different from that on classification tasks. The present study first time investigated the effects of model pruning/quantization on the SE task (a regression task), and the results can be used as a useful guidance for future SE studies.




\bibliographystyle{IEEEbib}



\end{document}